\documentclass[twoside]{ilcws08}
\usepackage[latin1]{inputenc}
\usepackage[dvips]{graphicx,epsfig,color}
\usepackage{wrapfig,rotating}
\usepackage{amssymb,amsmath,array}

\newcommand{\lhhh}{$\lambda_{HHH}$}
\newcommand{\eezhh}{$e^{+} e^{-} \to ZHH$}
\newcommand{\zhhnnhh}{$ZHH \to \nu \bar{\nu} HH$}
\newcommand{\zhhqqhh}{$ZHH \to q \bar{q} HH$}
\newcommand{\nnhh}{$\nu \bar{\nu} HH$}
\newcommand{\qqhh}{$q \bar{q} HH$}
\newcommand{\zhhqqhhnnhh}{$ZHH \to \nu \bar{\nu} HH/q \bar{q} HH$}

\begin{document}
\title{Analysis of Higgs Self-coupling with $ZHH$ at ILC} 
\author{Yosuke Takubo
\vspace{.3cm}\\
Department of Physics, Tohoku University, Sendai, Japan
}

\maketitle

\begin{abstract}
Measurement of the cross-section of \eezhh~offers the information of
the trilinear Higgs self-coupling,
which is important to confirm the mechanism of the electro-weak symmetry breaking.
Since there is huge background in the signal region,
background rejection is key point to identify $ZHH$ events.
In this paper, we study the possibility to observe the $ZHH$ events at ILC 
by using \zhhqqhhnnhh~events.
\end{abstract}

\section{Introduction}
In the standard model,
particle masses are generated through the Higgs mechanism.
This mechanism relies on a Higgs potential, 
$V(\Phi) = \lambda (\Phi^{2} - \frac{1}{2} v^{2})^{2}$,
where $\phi$ is an iso-doublet scalar field, and $v$  
is the vacuum expectation value of its neutral component ($v \sim 246$ GeV).
Determination of the Higgs boson mass,
which satisfies $m_{H}^{2} = 2 \lambda v^{2}$ at tree level in the standard model,
will provide an indirect information about the Higgs potential and its self-coupling, \lhhh.
The measurement of the trilinear self-coupling, 
$\lambda_{HHH} = 6 \lambda v$, offers 
an independent determination of the Higgs potential shape
and the most decisive test of the mechanism of the electro-weak symmetry breaking.

\begin{wrapfigure}{r}{0.5\columnwidth}
\centerline{\includegraphics[width=0.45\columnwidth]{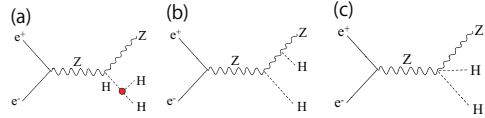}}
\caption{The relevant Feynman diagrams for the $ZHH$ production. 
The trilinear self-coupling is included in (a).}
\label{fig:diagram}
\end{wrapfigure}

\lhhh~can be extracted from the measurement of the cross-section 
for the Higgs-strahlung process ($\sigma_{ZHH}$), \eezhh.
For a Higgs mass of 120 GeV, the $W$ fusion process is negligible
at $\sqrt{s} = 500$ GeV.
Figure \ref{fig:diagram} shows the relevant Feynman diagrams for this process.
The information of \lhhh~is included in the diagram of Fig. \ref{fig:diagram}(a),
and the relation between the cross-section of 
$ZHH$ and \lhhh~is characterized by
$\frac{\Delta \lambda_{HHH}}{\lambda_{HHH}} 
\sim 1.75 \frac{\Delta \sigma_{ZHH}}{\sigma_{ZHH}}$,
where $\Delta \lambda_{HHH}$ and $\Delta \sigma_{ZHH}$
are measurement accuracy of $\lambda_{HHH}$ and $\sigma_{ZHH}$, respectively
\cite{castanier}.
For that reason, precise measurement of the cross-section for the $ZHH$ production
is essential to determination of the strength of the trilinear Higgs self-coupling.

We have studied the feasibility for observation of $ZHH$ events at the ILC.
For the analysis, we assumed a Higgs mass of 120 GeV, 
$\sqrt{s} = 500$ GeV, and an integrated luminosity of 2 $ab^{-1}$.
The final states of the $ZHH$ production can be categorized into 3 types, 
depending on the decay modes of $Z$:
$ZHH \to q\bar{q} HH$ (135.2 ab$^{-1}$), 
$ZHH \to \nu\bar{\nu} HH$ (38.8 ab$^{-1}$), and
$ZHH \to \ell\bar{\ell} HH$ (19.8 ab$^{-1}$),
where the cross-sections were calculated 
without the beam polarization, initial-state radiation, and beamstrahlung.
In this paper, we report status of the analysis with \zhhqqhhnnhh events.

\section{Simulation tools}
\begin{wrapfigure}{r}{0.35\columnwidth}
\centerline{\includegraphics[width=0.3\columnwidth]{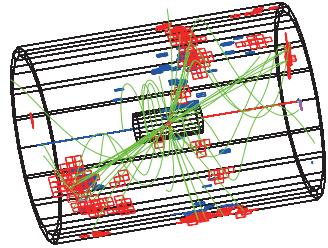}}
\caption{A typical event display of $ZHH \to \nu_{\mu} \bar{\nu}_{\mu} HH$.}
\label{fig:evtdsp}
\end{wrapfigure}

We have used MadGraph \cite{madgraph} 
to generate \zhhqqhhnnhh~and $tbtb$ events, 
where top quarks in $tbtb$ events are decayed by using DECAY package in MadGraph.
$ZZ \to bbbb$, $tt$, and $ZH$ events have been generated 
by Physsim \cite{physsim}.
In this study, the beam polarization, initial-state radiation, and beamstrahlung 
have not been included in the event generations. 
We also have ignored the finite crossing angle between the electron and positron beams. 
In both event generations, 
helicity amplitudes were calculated using the HELAS library \cite{helas}, 
which allows us to deal with the effect of gauge boson polarizations properly. 
Phase space integration and the generation of parton 4-momenta have been performed 
by BASES/SPRING \cite{bases}. 
Parton showering and hadronization have been carried out by using PYTHIA6.4 \cite{pythia}, 
where final-state tau leptons are decayed by TAUOLA \cite{tauola} 
in order to handle their polarizations correctly.

The generated Monte Carlo events have been passed to a detector simulator 
called JSFQuickSimulator, 
which implements the GLD geometry and other detector-performance 
related parameters \cite{glddod}. 
Figure \ref{fig:evtdsp} shows a typical event display of 
$ZHH \to \nu_{\mu} \bar{\nu}_{\mu} HH$.

\section{Analysis}

\subsection{\zhhnnhh} \label{sec:nnhh}
\begin{wrapfigure}{r}{0.38\columnwidth}
\centerline{\includegraphics[width=0.34\columnwidth]{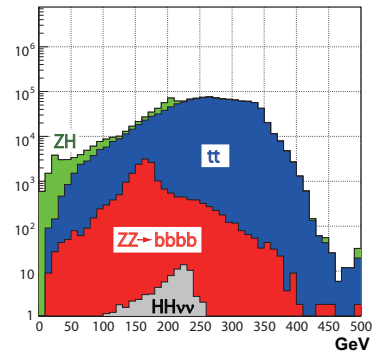}}
\caption{Distribution of the sum of the two reconstructed Higgs masses 
for \nnhh~and background events.}
\label{fig:nocut}
\end{wrapfigure}

For the Higgs mass of 120 GeV, the Higgs boson mainly decays into $b\bar{b}$ 
(76\% branching ratio in MadGraph).
Therefore, we concentrated on $ZHH \to \nu\bar{\nu} b\bar{b}b\bar{b}$
from \nnhh~events.
As background events, we considered $ZZ \to bbbb$ (9.05 fb), 
$tt$ (583.6 fb), $ZH$ (62.1 fb), and $tbtb$ (1.2 fb).
They have much larger cross-sections than $ZHH$,
necessitating powerful background rejection.

The clusters in the calorimeters are combined to form a jet 
if the two clusters satisfy $y_{ij} < y_{\mathrm{cut}}$,
where $y_{ij}$ is $y$-value of the two clusters. 
All events are forced to have four jets by adjusting $y_{\mathrm{cut}}$.
Then, mass of the Higgs boson was reconstructed to identify \nnhh~events
by minimizing $\chi^{2}$ value defined as
\begin{equation}
\chi^{2} = 
\frac{(^{\mathrm{rec}}M_{H1} - ^{\mathrm{true}}M_{H})^{2}}{\sigma_{H1}^{2}} +
\frac{(^{\mathrm{rec}}M_{H2} - ^{\mathrm{true}}M_{H})^{2}}{\sigma_{H2}^{2}},
\end{equation}
where $^{\mathrm{rec}}M_{H1,2}$, $^{\mathrm{true}}M_{H1,2}$, and
$\sigma_{H1,2}$ are the reconstructed Higgs mass, the true Higgs mass (120 GeV),
and the Higgs mass resolution, respectively.
$\sigma_{H1,2}$ was evaluated for each reconstructed Higgs boson
by using 31\%$/\sqrt{E_{\mathrm{jet}}}$, where $E_{\mathrm{jet}}$ is the jet energy.
Figure \ref{fig:nocut} shows the distribution of the sum of 
the two reconstructed Higgs boson masses 
for \nnhh~and background events.
With no selection cuts, 
the signal is swamped in huge number of background events.

To identify the signal events from the backgrounds, 
we applied the following selection cuts.
We required $\chi^{2} < 20$ and 95 GeV$< M_{H1,2} <$ 125 GeV
to select events, for which the Higgs bosons could be well reconstructed.
Since Higgs mainly decay into a b-quark pair,
the reconstructed mass distribution have a tail in lower mass region
due to missing energy by neutrinos from decay processes of the b-quark.
For that reason, the mass cut is applied asymmetrically 
against the Higgs mass.
Then, since a $Z$ boson is missing in \nnhh~events,
we set the selection cut on the missing mass ($^{\mathrm{miss}}M$):
90 GeV $< ^{\mathrm{miss}}M < $ 170 GeV.

\begin{wrapfigure}{r}{0.55\columnwidth}
\centerline{\includegraphics[width=0.5\columnwidth]{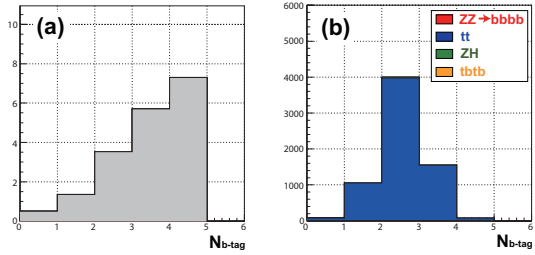}}
\caption{Distribution of the number of jets tagged as $b$-jets
after the selection cuts for \nnhh~(a) and backgrounds (b).}
\label{fig:nb}
\end{wrapfigure}

The angular distribution of the particles reconstructed as the Higgs bosons
has a peak at $\cos \theta = \pm 1$ for $ZZ$ events
whereas the distribution becomes more uniform in \nnhh~events. 
We applied the angular cut of $| \cos \theta_{\mathrm{H1,2}}| < 0.9$
to reject these $ZZ$ events.

The 4-jet events from $ZH$ events 
have small missing transverse momentum ($^{\mathrm{miss}} P_{\mathrm{T}}$),
which contaminate in the signal region.
For that reason, we required $^{\mathrm{miss}} P_{\mathrm{T}}$ above 50 GeV.

\begin{wrapfigure}{r}{0.38\columnwidth}
\centerline{\includegraphics[width=0.34\columnwidth]{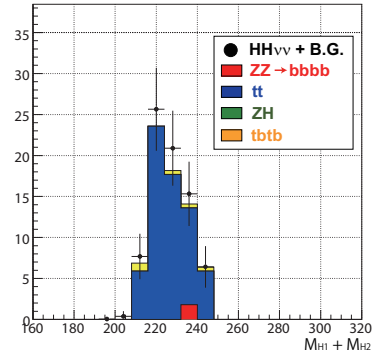}}
\caption{Distribution of the sum of the two reconstructed 
Higgs boson masses for \zhhnnhh~after all the selection cuts.}
\label{fig:bgsig}
\end{wrapfigure}

After the selection cuts so far, the dominant background was $tt$ events. 
The leptonic decay mode of $W$ from $t \to b W$ can be rejected 
by indentifying isolated charged leptons.
We define the energy deposit within 20 degree around a track as $E_{20}$.
The isolated lepton track was defined to be a track with 
10 GeV $< E_{20} < \frac{2}{11} E_{\mathrm{trk}} - 1.8$ GeV,
where $E_{\mathrm{trk}}$ is energy of the lepton track.
We required the number of isolated lepton tracks ($N_{\mathrm{lepton}}$) to be zero.

Finally, the flavor tagging was applied.
We identified a jet as a $b$-jet, when it has 2 tracks 
with 3-sigma separation from the interaction point.
Figure \ref{fig:nb} shows the distribution of the number of jets tagged as $b$-jets
after the selection cuts ($N_{\mathrm{b-tag}}$).
Since the Higgs boson decays into $b\bar{b}$ with a 76\% branching ratio,
\nnhh~events have a peak at $N_{b\mathrm{-tag}} = 4$,
whereas $tt$ events have a peak at 2.
To reject the $tt$ events effectively, we selected events with $N_{b\mathrm{-tag}} = 4$.

Figure \ref{fig:bgsig} shows the distribution of 
the sum of the two reconstructed Higgs masses for \zhhnnhh~after all the selection cuts.
We summarize the reduction rate by each selection cut in Table \ref{tb:cut_rate}.
Finally, we obtained 7.3 events for \nnhh~and 69.2 events for backgrounds.
This result corresponds to a signal significance of 0.8 ($= 7.3/\sqrt{7.3 + 69.2}$).
For observation of the $ZHH$ production, further background rejection, 
especially $tt$ events, is necessary.

\begin{table}[t]
\centerline{
\begin{tabular}{|l|rrrrr|} \hline
                & \nnhh       & $ZZ \to bbbb$ & $tt$      & $ZH$    & $tbtb$  \\ \hline
No cut          & 77.6        & 18,100        & 1,167,200 & 124,200 & 2,154 \\
$\chi^{2} < 20$ & 43.7        & 12,169        & 364,921   & 83,065  & 468 \\
95 GeV$< M_{H1,2} <$ 125 GeV & 29.5 & 387 & 70,557 & 8,570 & 82 \\ 
90 GeV$< ^{\mathrm{miss}}M <$ 170 GeV & 26.2 & 127 & 32,570 & 696  & 45 \\
$| \cos \theta_{H1,2}| < 0.9$ & 23.0 & 34.4 & 26,521 & 447 & 37 \\
$^{\mathrm{miss}} P_{\mathrm{T}} > 50$ GeV & 18.4 & 3.6 & 17,591 & 137 & 25 \\
$N_{\mathrm{lepton}} = 0$ & 17.8 & 3.6 & 6,708 & 37.3 & 9.7 \\
$N_{\mathrm{b-tag}} = 4$ & 7.3 & 1.8 & 65 & 0 & 2.4 \\   
\hline
\end{tabular}}
\caption{Cut statistics.}
\label{tb:cut_rate}
\end{table}

\subsection{\zhhqqhh}
For the analysis of $qqHH$, 
all the events are reconstructed as 6-jet events, adjusting the $y$-value.
Here, we considered $tt$ and $tbtb$ events as background events.
The masses of the Higgs and Z boson were reconstructed 
by minimizing $\chi^{2}$ value defined as
\begin{equation}
\chi^{2} = 
\frac{(^{\mathrm{rec}}M_{H1} - ^{\mathrm{true}}M_{H})^{2}}{\sigma_{H1}^{2}} +
\frac{(^{\mathrm{rec}}M_{H2} - ^{\mathrm{true}}M_{H})^{2}}{\sigma_{H2}^{2}} +
\frac{(^{\mathrm{rec}}M_{Z} - ^{\mathrm{true}}M_{Z})^{2}}{\sigma_{Z}^{2}},
\end{equation}
where $^{\mathrm{rec}}M_{H1,2}$, $^{\mathrm{rec}}M_{Z}$,
$^{\mathrm{true}}M_{H1,2}$, and $^{\mathrm{true}}M_{Z}$ are
the reconstructed Higgs and Z mass and the true Higgs and Z mass, respectively.
$\sigma_{H1,2}$ and $\sigma_{Z}$ are the Higgs and Z mass resolution, respectively, 
which are defined in Sec \ref{sec:nnhh}.

\begin{wrapfigure}{r}{0.38\columnwidth}
\centerline{\includegraphics[width=0.34\columnwidth]{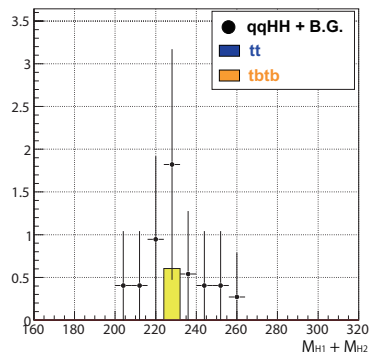}}
\caption{Distribution of the sum of the two reconstructed 
Higgs boson masses for \zhhqqhh~after all the selection cuts.}
\label{fig:bgsig}
\end{wrapfigure}

We required $\chi^{2} < 20$, 90 GeV$< M_{H1,2} <$ 150 GeV,
and 60 GeV$< M_{Z} <$ 120 GeV to select events, 
for which the Higgs and Z bosons could be well reconstructed.
Then, the isolated lepton track was searched to indentify the lepton tracks 
from decay of top quarks in $tt$ and $tbtb$ events.
We required the number of isolated lepton tracks ($N_{\mathrm{lepton}}$) to be zero.
Since the missing energy of the signal is smaller than $tt$ and $tbtb$ events, 
$^{\mathrm{miss}} E < 70$ GeV was required.
Finally, we applied the b-tagging whose requirement is the same as the analysis for \nnhh events.
Here, we required that all the jets are b-jets, $N_{\mathrm{b-tag}} = 6$.

After all the cut, 
we obtained 4.6 events for $qqHH$ and 0.6 events for the background.
That corresponds to the signal significance of 2.0 ($= 4.6/\sqrt{4.6 + 0.6}$).
The number of the events at each selection cut is summarized 
in Table \ref{tb:cut_rate}.

\begin{table}[tb]
\centerline{
\begin{tabular}{|l|rrr|} \hline
                & $qqHH$    & $tt$      & $tbtb$  \\ \hline
No cut          & 270       & 1,167,200 & 124,200 \\
$\chi^{2} < 20$ & 219       & 615,456   & 1,810    \\
90 GeV$< M_{H1,2} <$ 150 GeV  & 214 & 600,899 & 1,781 \\ 
60 GeV$< M_{Z} <$ 120 GeV     & 213 & 595,533 & 1,771 \\
$N_{\mathrm{lepton}} = 0$     & 193 & 467,154 & 1,240 \\
$^{\mathrm{miss}} E < 70$ GeV & 170 & 352,061 & 943 \\
$N_{\mathrm{b-tag}} = 6$      & 4.6 & 0       & 0.6 \\   
\hline
\end{tabular}}
\caption{Cut statistics.}
\label{tb:cut_rate}
\end{table}

\section{Summary}
\zhhqqhhnnhh~processes were studied to investigate 
the possibility of the trilinear Higgs self-coupling at the ILC.
In this study, we assumed the Higgs boson mass of 120 GeV,
$\sqrt{s} = 500$ GeV, and the integrated luminosity of 2 ab$^{-1}$.
After the selection cuts,
the signal significance of 0.8 and 2.0 was obtained
for \nnhh~and \qqhh events, respectively.
To extract the information of \lhhh,
we must improve the flavor tagging to reject background events effectively.

\section{Acknowledgments}
The authors would like to thank all the members of the ILC physics subgroup
\cite{softg} for useful discussions. 
This study is supported in part by the Creative Scientific Research Grant
No. 18GS0202 of the Japan Society for Promotion of Science, and
Dean's Grant for Exploratory Research in Graduate School of Science of Tohoku University.


\begin{footnotesize}




\begin{thebibliography}{99}

\bibitem{castanier} C. Castanier, P. Gay, P. Lutz, J Orloff, arXive:hep-ex/0101028.
\bibitem{madgraph} http://madgraph.hep.uiuc.edu/.
\bibitem{physsim} http://acfahep.kek.jp/subg/sim/softs.html.
\bibitem{helas} H. Murayama, I. Watanabe, K. Hagiwara, KEK-91-11, (1992) 184.
\bibitem{bases} T. Ishikawa, T. Kaneko, K. Kato, S. Kawabata,\emph{Comp, Phys. Comm.} {\bf 41} (1986) 127.
\bibitem{pythia} T. Sj$\dot{\mathrm{o}}$strand, \emph{Comp, Phys. Comm.} {\bf 82} (1994) 74.
\bibitem{tauola} http://wasm.home.cern.ch/wasm/goodies.html.
\bibitem{glddod} GLD Detector Outline Document, arXiv:physics/0607154.
\bibitem{softg} http://www-jlc.kek.jp/subg/physics/ilcphys/.
\end{thebibliography}
%

\end{footnotesize}


\end{document}